# Light assisted collisions in ultra-cold Tm atoms

I.S. Cojocaru[1,2], S.V. Pyatchenkov[1], S.A. Snigirev[1], I.A. Luchnikov[1,2], E.S. Kalganova[1,2,4], G.A.Vishnyakova[1,4], D.N. Kublikova[1,2] V.S. Bushmakin[1,2], E.T. Davletov[1,2], V.V. Tsyganok[1,2], O.V. Belyaeva[1], A. Khoroshilov[1], V.N. Sorokin[1,4], D.D. Sukachev[1,4,5], A.V. Akimov[1,3,4]

[1]Russian Quantum Center, Business center "Ural", 100A Novaya str., Skolkovo, Moscow, 143025, Russia

[2]Moscow Institute of Physics and Technology, Institutskii per. 9, Dolgoprudny, Moscow region, 141700, Russia

[3]Texas A&M University, 4242 TAMU, College Station, Texas, 77843, USA,

[4]PN Lebedev Institute RAS, Leninsky prospekt 53, Moscow, 119991, Russia,

[5]Harvard University, Physics Department, 17 Oxford str., Cambridge, MA, 02138 USA

e-mail: akimov@physics.tamu.edu

## I.   ABSTRACT

We studied light assisted collisions of Tm atoms in a magneto optical trap (MOT) for the first time, working on a weak cooling transition at 530.7 nm ( $4f^{13}(^2F^o)6s^2, J=7/2, F=4$ to $4f^{12}(^3H_6)5d_{5/2}6s^2, J=9/2, F=5$ ). We observed a strong influence from radiation trapping and light assisted collisions on the dynamics of this trap. We carefully separated these two contributions and measured the binary loss rate constant at different laser powers and detuning frequencies near the cooling transition. Analyzing losses from the MOT, we found the light assisted inelastic binary loss rate constant to reach values of up to $\beta = 10^{-9}$ cm$^3/$s and gave the upper bound on a branching ratio $k < 0.8 \times 10^{-6}$ for the 530.7 nm transition.

## II.   INTRODUCTION

Laser cooling and trapping of neutral atoms has become a powerful tool enabling a number of research activities and applications such as quantum simulations [1–3], quantum information processing [4], control of light propagation [5,6], optical pulse switching [7], metrology [8,9], and testing the drift of the fundamental constants [10,11]. Recently, rare earth elements have attracted considerable attention due to high orbital and magnetic momenta [12,13]. Low field magnetic Feshbach resonances and large anisotropic dipole-dipole interaction [14,15] in lanthanides give rise to number of effects making physics of degenerate dipolar gases very attractive [13,16,17].

The only stable isotope of Thulium — $^{169}$Tm is bosonic and has one hole in the inner 4f electronic shell shielded by closed outer $5s^2$ and $6s^2$ shells. The ground state of $^{169}$Tm has a total angular momentum

$F=4$ and a magnetic momentum of $4\mu_B$, where $\mu_B$ is Bohr magneton. While magnetic moment of the Tm atom is less than that of Er and Dy [12,13], a relatively simple level structure and the possibility to capture Tm atoms in a 532 nm dipole trap [18,19] make the Tm atom an attractive species for quantum simulations. On the other hand, a magneto-dipole transition between fine structure sublevels of the Tm ground state at 1.14 $\mu$m has a linewidth less than 1.4 Hz and is promising for optical clocks applications [11,20]. Laser cooling down to 25 µK [21], trapping into a narrow line magneto optical trap (MOT), and loading into magnetic and various optical dipole traps have been recently demonstrated [22,23].

Collisions in ultracold quantum gases play a significant role in metrology and quantum simulations [24–26]. For example, understanding of collisional properties enables control over particles interaction and the optimization of cooling schemes, especially evaporative [27] and demagnetization cooling [28].

Collisional properties of Tm atoms were studied in several papers over the last few decades [29–31]. Buffer gas cooling enabled measurement of Tm-He and Tm-Tm dipolar relaxation rates at around 1 mK temperatures [23,29]. Light assisted collisions of thulium atoms have not been investigated systematically yet (in [32] authors estimated the binary collisions loss rate constant only for fixed intensity and frequency detuning), even though they do play a significant role in limiting the maximum density of atoms in the trap. Light assisted collisions lead to a trap loss either due to a radiative escape (RE) or a hyperfine or fine structure changing collision, a relative contribution which depends on laser detuning and other parameters [33–36]. Typical values of the light assisted binary collisions rate constants are in the range of $\beta \sim 10^{-13} \div 10^{-9}\,\text{cm}^3/\text{s}$ for small (below 10 linewidth of the cooling transition) laser detunings and deep (around 0.5 mK) traps. For shallow microtraps, collisional rate constants can exceed $\beta > 10^{-8}\,\text{cm}^3/\text{s}$ [37]. In this paper, we demonstrated strong inelastic rate constants exceeding $10^{-9}\,\text{cm}^3/\text{s}$ in case of a deep, not far detuned trap.

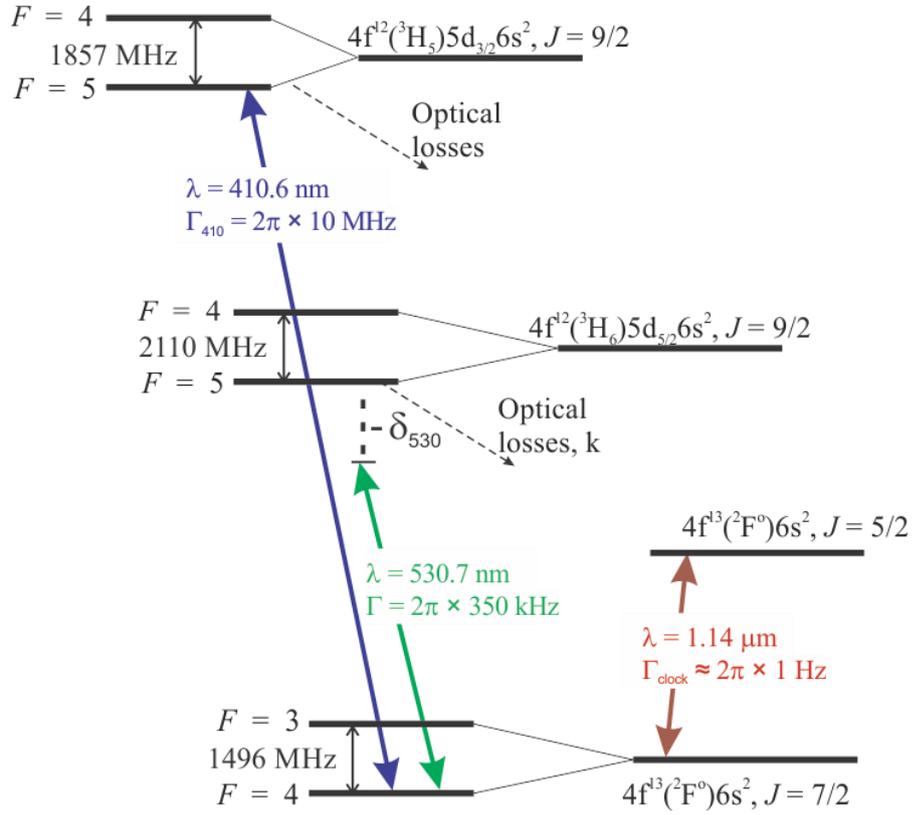

Figure 1. Relevant energy levels of Tm atom. Both $4f^{12}(^3H_6)5d_{5/2}6s^2, J=9/2, F=5$ and $4f^{12}(^3H_5)5d_{3/2}6s^2, J=9/2, F=5$ levels have weak decay channels to the level $4f^{13}(^2F^o)6s^2, J=5/2$ and other levels not shown at this figure.

## III. SETUP LAYOUT

Tm levels involved in our experiment are shown in Figure 1. The Zeeman slower (ZS) [38], the 3D optical molasses [39,40], and absorption imaging used a strong almost cycling 410.6 nm transition ($F=4\rightarrow F=5$) with natural linewidth of $2\pi\times 10$ MHz. A weak 530.7 nm transition ($F=4\rightarrow F=5$) with a natural linewidth of $\Gamma=2\pi\times 350$ kHz and a corresponding Doppler temperature of 9 $\mu$K was used for the single stage MOT. This transition has a small branching ratio (see Section IV.C) and there is no need of the repumping laser for the MOT [22]. Atoms were loaded in the MOT directly from the ZS. This approach allows to have a higher number of trapped atoms by eliminating the short lived MOT working on the 410.6 nm transition [41].

Figure 2 represents a scheme of an experimental setup. A thulium atomic beam was formed by an effusion cell (CreatecSFC-40 2-HL) operating at a temperature between $600\,°C$ and $800\,°C$. The 80 centimeter long ZS operating at the 410.6 nm transition in a "spin flip" configuration [42] decelerated hot atoms down to 20 m/s. The ZS laser beam had a frequency detuning of $-230$ MHz from the 410.6 nm transition (negative detuning means laser frequency smaller than resonant one), power of 30 mW, and a radius of 1.6 mm at

$1/e$ level of maximum intensity. A magnetic field profile is shown in Figure 2. The ZS was separated from the effusion cell by a pneumatically controlled valve, which could completely block the atomic beam.

The MOT in retro reflected configuration [43] worked on the 530.7 nm transition. A radius of the MOT beams on $1/e$ intensity level was 6.9 mm, and the intensity $I$ of the each beam varied from $10 I_s^{530}$ to $55 I_s^{530}$ (where $I_s^{530} = 3.16 \, \text{W/m}^2$ is the saturation intensity of the 530.7 nm cooling transition). The frequency detuning of the cooling laser from the 530.7 nm cooling transition $\delta_{530}$ was controlled by an acousto optic modulator (Gooch&Housego model number 350-192) and varied from $-0.2$ to $-0.3 \, \text{MHz}$ (see Figure 2). Magnetic field gradients were produced by two coils with anti-parallel currents and set to 6.5, 3.2, and 3.2 G/cm, for $z$ (along gravity), x, and y axis, respectively. An ion pump (Gamma Vacuum 300TV) evacuated the main vacuum chamber to a pressure below $5 \cdot 10^{-10} \, \text{mBar}$.

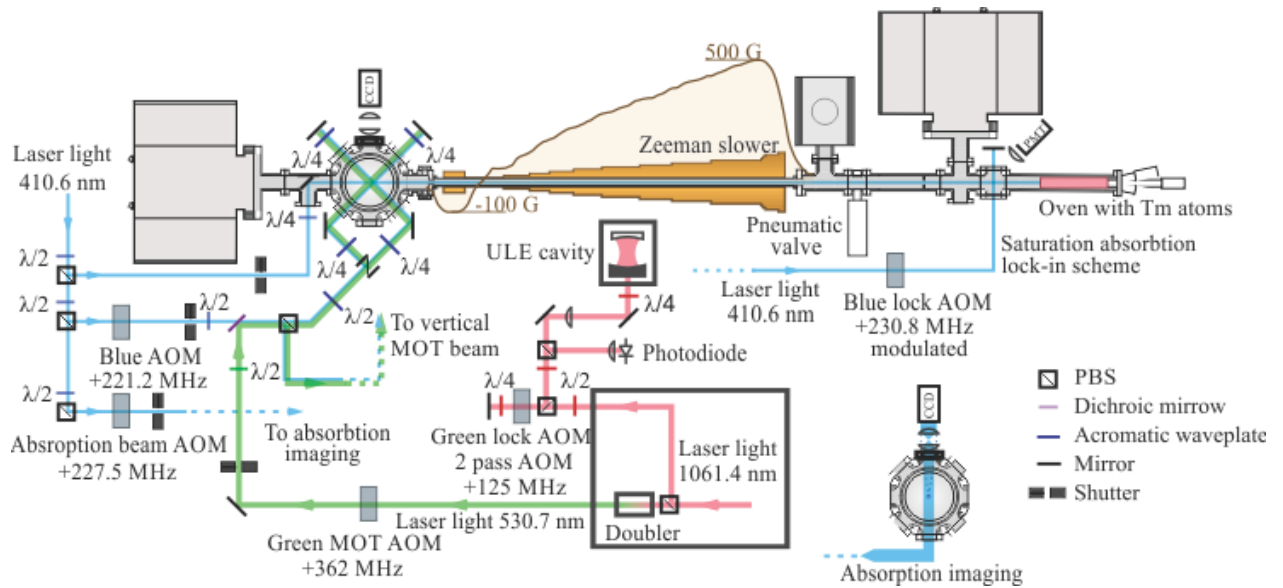

**Figure 2.** Experimental setup. PBS stays for polarization beam splitter, Blue AOM denotes acousto optic modulator (AOM) controlling a 3D optical molasses, Absorption beam AOM stands for AOM controlling frequency detuning of an imaging beam, Green lock AOM depicts AOM controlling frequency detuning of 530.7 nm laser from the reference cavity (ULE cavity), Blue lock AOM stands for AOM controlling frequency detuning of 410.6 nm laser from $F = 4 \rightarrow F = 5$ atomic transition at 410.6 nm. The measured Zeeman slower magnetic field is overlaid with the Zeeman slower itself. A blue molasses beam is shifted with respect to a MOT beam by 11 mm (center-to-center distance) towards the Zeeman slower. Note, that a MOT beam radius is 6.9 mm, so visually beams nearly touch each other. Both beams use the same mirror set.

The MOT capture velocity was about 5 m/s and was significantly lower than the ZS output velocity (20 m/s). A relatively large distance (20 cm) between ZS output and a MOT region made it difficult to reduce the ZS output velocity because it would cause a large divergence of the atomic beam due to transverse heating in the ZS. To decelerate atoms below the capture velocity and reduce the divergence of the atomic beam, we implemented an addition 3D optical molasses after the ZS, operating at the 410.6 nm transition. This increased the number of trapped atoms $N$ by factor of 6. The 3D molasses was located 11 mm before the center of the MOT region, each beam had a radius of 2.5 mm, a peak intensity of $I = 1.5 I_s^{410}$ (where $I_s^{410} = 180$ W/m² is the saturation intensity of the 410.6 nm transition), and a detuning of $\delta_{410} = -17$ MHz from the 410.6 nm transition.

Atoms in the MOT were detected by using the absorption imaging technique [44] (APPENDIX A). An atomic cloud was illuminated by an imaging laser beam resonant with the strong 410.6 nm transition. A transmitted beam was detected with a CCD camera (Thorlabs DCU-223M). The imaging beam was much wider than the atomic cloud and had intensity of $0.3 I_s^{410}$ and circular polarization. The temperature of the atoms in the cloud was measured by a time of flight technique [45] after switching off all laser beams and magnetic fields.

A 410 nm laser (Toptica TA-SHGpro) frequency was stabilized within 1.7 MHz near the 410.6 nm transition using the saturated absorption scheme [46] in the atomic beam right after the effusion cell. A laser operating at 530.7 nm (Toptica TA-SHGpro) was locked to an ultra low expansion (ULE) cavity (finesse of $10^5$, Stable Laser Systems) by Pound–Drever–Hall's scheme [47] and had a linewidth less than 100 kHz (measured at 0.1 s window, see APPENDIX B).

## IV. EXPERIMENT

We used the following sequence to measure the binary collisions loss rate constant as a function of laser beam intensity $I$ and frequency detuning $\delta_{530}$. First atoms were prepared at a temperature of 70 μK in a cloud of approximately Gaussian shape with a radius $w = 550$ μm measured on $1/e$ level (trap is loaded for 2.5 s with $\delta_{530} = -5.3$ MHz and $I = 63 I_s^{530}$ per beam). Then the ZS and the 3D optical molasses laser beams were blocked with shutters, the atomic beam was blocked with the pneumatic valve (at this point the number of atoms in the MOT started to decay), and MOT parameters were ramped to the desired values. The atoms equilibrated within the new parameters in 0.2 s. After that, the number of remained atoms in the cloud and the cloud radius were measured as a function of elapsed time. Magnetic field gradient of the MOT was on all the time (for pulse sequence, see APPENDIX C).

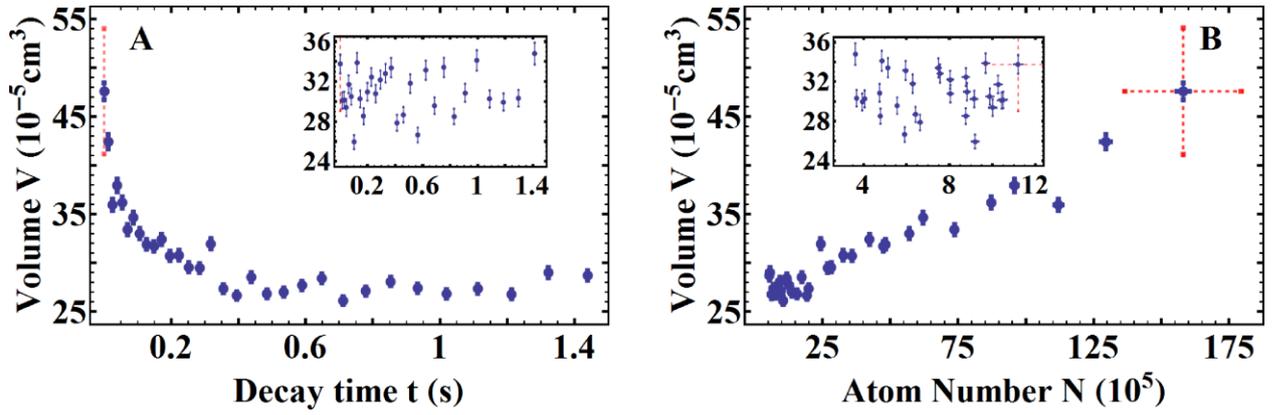

Figure 3 Volume of atomic cloud. A) Volume of the cloud captured with following parameters: $\delta_{530} = -1.8$ MHz, power per beam $P = 7.5$ mW, and initial number of atoms in the MOT $N = 2 \cdot 10^7$ (for details, see APPENDIX A). Inset shows volume with $\delta_{530} = -1.65$ MHz and $N = 10^6$ (axes are the same). B) Volume of the cloud versus number of trapped atoms, $\delta_{530} = -1.8$ MHz, $P = 7.5$ mW. Inset represent more detailed picture for small number of atoms in the trap with $\delta_{530} = -1.65$ MHz. Red dashed and blue solid error bars indicate full error and standard deviation (only fit error) correspondently, for details see APPENDIX D.

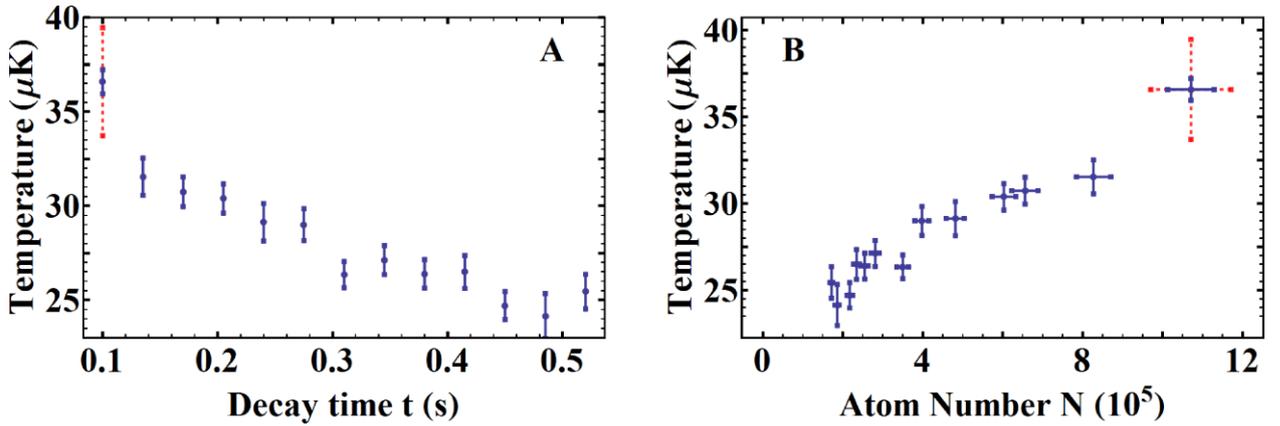

Figure 4 Temperature of the atomic cloud captured with $\delta_{530} = -2.6$ MHz and power per beam of 26 mW. A) Temperature versus decay time after turning off the loading beam of thulium atoms. B) Temperature versus the number of atoms in the MOT. Red dashed and blue solid error bars indicate full and statistical error, correspondently, for details see APPENDIX D.

### A. Radiation trapping

We have found that the atomic cloud volume $V$ depended on the decay time (see Figure 3A). A possible reason for that is the radiation trapping (RT) that takes place in MOTs with a relatively high optical depth [48,49]. In this case, reabsorption of a photon scattered by another atom in the MOT leads to a repulsion force between atoms. This effect limits the maximum possible density in the center of the atomic cloud and makes the cloud volume somehow proportional to the number of atoms $N(t)$. To prove this, we measured $V$ as a function of $N$ (Figure 3B), varying the flux of slowed atoms after the ZS and keeping all other parameters constant. When the initial number of trapped atoms is small, the RT is not expected and volume did not depend on the decay time (Figure 3, insets). At a large number of atoms (the maximum resonance optical depth in these measurements is $\approx 1.5$), the RT takes place.

Correspondently, the volume began to depend on decay time (Figure 3A). Furthermore, due to a photon recoil energy, the reabsorption process increases the temperature of the cloud. It follows then that the temperature should decrease with the decay of the MOT population. Indeed, we observed this behavior in our MOT (see Figure 4). Therefore, we concluded that the RT is a mechanism responsible for the variation of the volume in our trap.

## B. Collisions in the MOT

The decay of the number of the atoms in the MOT after shutting off the atomic beam and 3D optical molasses is governed by the equation [24]:

$$\frac{dN(t)}{dt} = -\gamma N(t) - \beta \frac{N^2(t)}{(2\pi)^{3/2} w^3(t)}, \quad (1)$$

where $\gamma$ is a linear loss rate, $\beta$ is an inelastic binary collision loss rate constant, and $w$ is the radius at level $1/e$ of an atomic cloud which has a Gaussian density profile. Therefore, we can deduce the loss rates $\gamma$ and $\beta$ from an analysis of the decay of the number of trapped atoms [24]. The typical decay curves are shown in Figure 5A. When atomic concentration is large (at short decay time) the second term on the right side of (1) dominates. At a longer times $N$ is reduced and the first term governs the decay process.

## C. Branching ratio of the cooling transition.

Collisions with the residual gas in the vacuum chamber and the small non-cyclicity of the 530.7 nm cooling transition contribute to the linear loss rate $\gamma$. In a low-density limit, (1) has a simple exponential solution:

$$\begin{aligned} N(t) &= N_0 \exp(-\gamma t) \\ \gamma &= \gamma_0 + \gamma_1 = \gamma_0 + k\rho_{ee}\Gamma \end{aligned} \quad (2)$$

where $\gamma_0$ is the loss rate associated with the collisions of the trapped atoms with the residual gas in the vacuum chamber, $\gamma_1$ is the loss rate due to spontaneous decay of the upper cooling level to other states, $k$ is the corresponding branching ratio, and $\rho_{ee}$ is the population of the upper cooling level. To measure the number of atoms in the MOT, we also collected the resonance fluorescence from the same 410 nm imaging beam because it gave a higher signal to noise ratio for long time delays than the absorption imaging. Then the tails of the time dependence on number of atoms (where binary collisions are negligible) was fitted by (2) and the value for the decay rate $\gamma$ was determined for each detuning.

To find a branching ratio k of the upper level of 530.7 nm cooling transition, the population of this level during the MOT operation for certain $\delta_{530}$ and $I$ has to be found.

Since the dominating process in the MOT is optical pumping, we ignored any atomic coherence and used rate equations for each magnetic sublevels of excited and ground states in the presence of spatially varying MOT quadrupole magnetic field $B$:

$$\sum_{v=\pm 1, 0} \left(C^{v}_{m_e, m_g}\right)^2 \left(S_v \left(\rho_{m_e} - \rho_{m_g}\right) + \rho_{m_e}\right) = 0, \quad m_g = -F, \ldots, F \text{ and } m_e = v + m_g,$$

$$S_v = s / \left(1 + 4\left(2\pi\delta_{530} + v\mu_B B/\hbar\right)^2 / \Gamma^2\right)$$

(3)

where $s = I/I_s^{530}$ is the saturation parameter per MOT beam, $C^{v}_{m_e, m_g}$ is the Clebsch-Gordan coefficient, $v$ represents light polarization ($v=+1$ corresponds to $\sigma^+$ polarization, $v=-1$ to $\sigma^-$ and 0 to $\pi$ with respect to z-axis) $\rho_{m_e}, \rho_{m_g}$ are the populations of exited and ground state magnetic sublevels, correspondently, and $\hbar$ is the reduced Plank's constant. In this equation, we used a fact that intensities of all light polarizations in a MOT are identical.

Then we averaged total population of the excited state over the atomic cloud spatial distribution,

$$\rho_{ee} = \left\langle \sum_{m_e} \rho_{m_e} \right\rangle.$$

(4)

We estimated the branching ratio from the slope of the calculated $\gamma(\rho_{ee})$ dependence (Figure 5B). Linear fit gave us an upper bound of 530.7 nm transition non-cyclisty at the level of $k = 0.7 \times 10^{-6}$ with a 95% tolerance interval up to $k = 0.8 \times 10^{-6}$. Unfortunately, dependence of loss rate on fraction of the atoms in the exited state may also be explained by assuming presence of a non-vanishing contribution of light assisted collisions even at large time delays rather than assuming finite branching ratio of the existed state. Since we do not have any clear way to measure or model to calculate a possible contribution of light assisted collisions to the curve on Figure 5B, we cannot estimate lower bound of the branching ratio. Indeed, any contribution of binary collisions in the tails of MOT decay data will lead to an overestimated branching ratio, therefore we gave the bound for the branching ratio as $k < 0.8 \times 10^{-6}$.

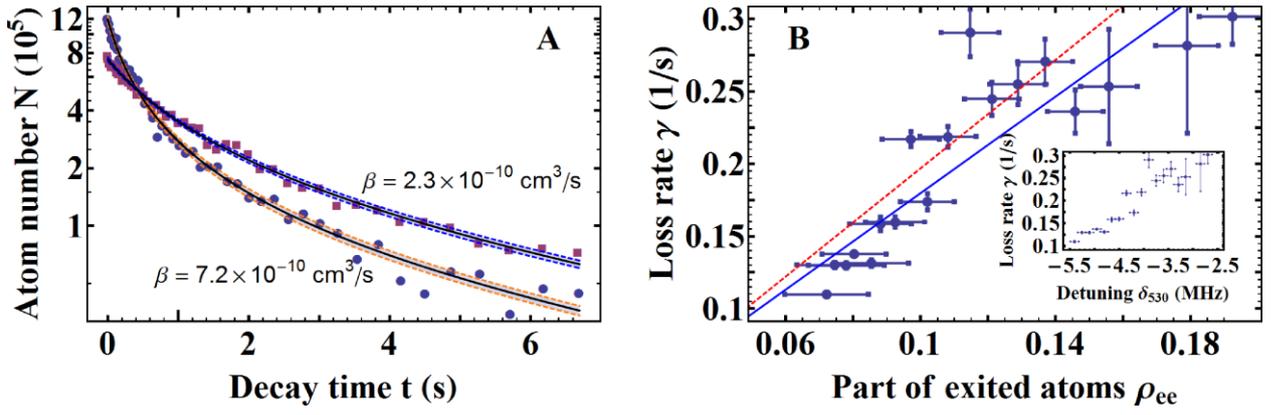

Figure 5 A) Number of atoms in the MOT during its decay. Solid lines represent fit, dashed lines are standard deviation from the fitted curve. Curves were taken at detunings of −3.4 MHz (squares) and −2 MHz (circles) and cooling laser power per beam of 26 mW. B) The linear loss rate as a function of the upper state population $\rho_{ee}$ which was calculated from the known detuning and power of the cooling beams. Lines represent linear fit (blue) and upper bound with 95% confidence level (dashed red line). Inset represents original dependence of the linear loss rate on the detuning taken at MOT beam power of 26 mW per beam.

### D. Binary collisions.

Light assisted inelastic collisions comprise radiative escape and hyperfine or fine structure changing collisions. The first process occurs due to a resonant dipole-dipole interaction, in which atoms gain kinetic energy sufficient to leave a trap. In the second case, after a collision atoms are transferred to an untrapped state [24,33–36]. These collisions lead to the non-exponential character of atomic cloud decay (Figure 5A). To minimize the influence of the RT and to make the atomic cloud radius constant during the measurements (see Figure 3A), we adjusted the temperature of the effusion cell to capture less than $2 \times 10^6$ atoms. In this case, (1) has a following solution:

$$N(t) = N_0 \frac{e^{-t\gamma}}{1 + N_0 \frac{\beta}{(2\pi)^{3/2} w^3 \gamma}\left(1 - e^{-t\gamma}\right)}, \qquad (5)$$

which is valid for a Gaussian spatial distribution of atoms in the atomic cloud. First, we fitted (5) to the data by varying both $\gamma$ and $\beta$. This led to large fit residuals and large errors for the binary loss rate constant $\beta$. Then, we tried to understand a mutual influence of parameters $\gamma$ and $\beta$ in the fitting procedure. We repeated the fitting procedure several times with a different *fixed* parameter $\gamma$ in the range of measured values of $\gamma$ (0.1–0.3 s$^{-1}$, see Figure 5B) and found the parameters generating the smallest fit residuals. A corresponding value of the parameter $\beta$ differed by less than 10% from the value obtained by using the smallest measured value of $\gamma = .12$ s$^{-1}$ (Figure 5A). As it was mentioned in the previous section, assumption of constant $\gamma$ (which can be explained as collisions of trapped atoms with buffer gas) does not contradict our data and, moreover, fits experimental data quite consistently. Fitting with non fixed $\gamma$ in most of the cases gave the same result for $\beta$ (within 10%). To eliminate an uncertainty associated with a choice of $\gamma$, we used a fixed value of $\gamma = .12$ s$^{-1}$ in the rest of this section. The error in $\beta$ due to a deviation of the actual distribution from the Gaussian one was estimated to be ~20% (see APPENDIX D). Taking into account a 25% error of measured atomic cloud radius and parameters fluctuations (see APPENDIX D), total uncertainty of the binary loss rate constant $\beta$ was less than 35%.

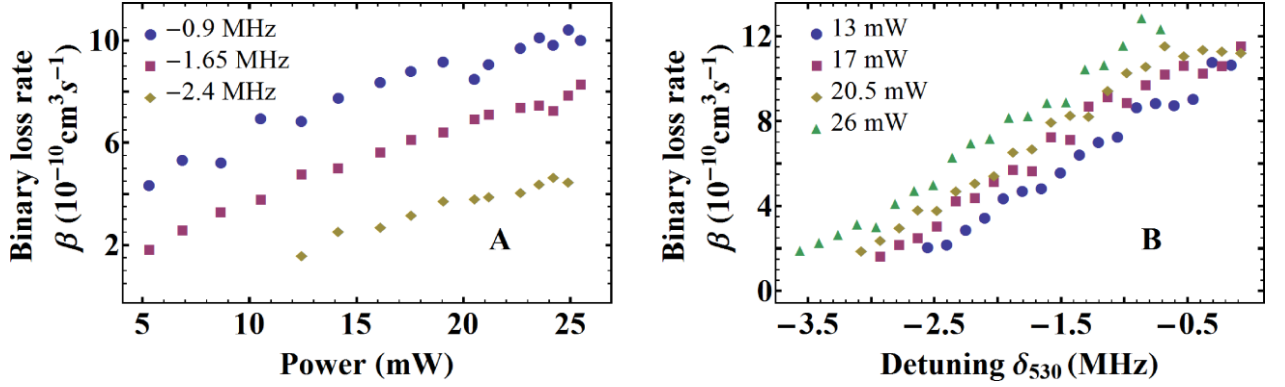

Figure 6. A) Binary loss rate constant $\beta$ versus power per MOT beam for different detunings $\delta_{530}$. B) Binary loss rate constant $\beta$ for different laser power per beam versus detuning $\delta_{530}$. The uncertainty of the binary loss rate constant does not exceed 35%, uncertainty of the detuning is no more than $-0.2$ MHz, and power uncertainty is estimated to be 5%

Finally, we measured the dependence of binary collision loss rate on the MOT cooling light detuning $\delta_{530}$ (Figure 6A) and the power $P$ (Figure 6B). Unfortunately, none of the simple analytical models such as Gallagher-Pritchard model [33], Julienne-Vigué model [50] and an improved model of radiative escape [51] describes the observed behavior well. For Tm atoms, these models can't explain the relatively large binary loss rate constant, which exceeds $10^{-9}$ cm$^3$/s for small light detuning. This large value of the loss rate for very deep traps makes us believe that fine or hyperfine structure changing collisions are more probable then the radiative escape. A detailed understanding of this large binary loss rate constant and its dependences on laser light parameters requires accurate quantum mechanical calculations [24,35,50]. Interaction potentials between thulium atoms have not been measured and such calculations are not yet possible.

## V. CONCLUSION

We studied light assisted collisions in Tm atom near the narrow 530.7 nm cooling transition in the MOT. Strong dependence of the rate constant on the trap parameters verified the light assisted nature of these collisions [50]. We measured a large binary collision rate constant $\beta > 10^{-9}$ cm$^3$/s. Simple analytical models failed to calculate such large rate constants. We also observed the radiation trapping in the MOT cloud, which caused the dependence of the MOT volume and temperature on the number of trapped atoms. Finally, we found the bound on the branching ratio $k$ of the upper level of the 530.7 nm transition ($4f^{12}(^3H_6)5d_{5/2}6s^2, J = 9/2, F = 5$) towards the levels not involved in cooling scheme to be $k < 0.8 \times 10^{-6}$.

## VI.　ACKNOWLEDGMENTS

This work was supported by Russian Science Foundation grant # 17-12-01419.

## APPENDIX A: ABSORPTION IMAGING AND FITTING PROCEDURE

For imaging, we used circularly polarized light which pumped all atoms onto a magnetic sublevel with $m = F$ (see APPENDIX D). In this case, intensity of the imaging beam after passing through the MOT cloud is the solution of the following equation at $z \to \infty$,

$$\frac{dI(x,y,z)}{dz} = -\frac{\sigma_0}{1 + I(x,y,z)/I_s} I(x,y,z) n(x,y,z)$$

$$\sigma_0 = \frac{3\lambda^2}{2\pi},$$

(6)

where $\sigma_0$ is a resonant absorption cross section of a single atom at wavelength $\lambda$ (in our case, $\lambda = 410.6$ nm), $n(x,y,z)$ is a density of atoms in the MOT at a point with coordinates $(x,y,z)$, and $x, y$ are coordinates transverse to the imaging beam propagation direction $z$. To measure the cloud parameters, two sequential absorption images were taken: one with atoms $I_c(x, y)$ and another one without atoms $I_0(x, y)$. Then we calculated a renormalized distribution $f(x, y)$ representing a density of atoms integrated along the direction of the imaging beam:

$$f(x,y) = \ln\frac{I_c(x,y)}{I_0(x,y)} + \frac{I_c(x,y) - I_0(x,y)}{I_s} = \sigma_0 \int n(x,y,z) dz.$$

The resulting distribution $f(x, y)$ was integrated over two orthogonal axes to get 1D density distributions $\sigma_0 n_x(x)$ and $\sigma_0 n_y(y)$, where

$$n_x(x) = \iint n(x,y,z) dy dz$$
$$n_y(y) = \iint n(x,y,z) dx dz.$$

The 1D densities were then fitted to a Gaussian function with parameters $A_i, B_i, x_0, y_0, w_x,$ and $w_y$:

$$\sigma_0 n_x(x) = B_x + A_x e^{-\left(\frac{x-x_0}{w_x}\right)^2}$$

$$\sigma_0 n_y(y) = B_y + A_y e^{-\left(\frac{y-y_0}{w_y}\right)^2}$$

The results of this fit were used to extract the number of atoms as:

$$N = \sqrt{\pi A_x A_y w_x w_y}.$$

The effective volume $V$ of the cloud was approximated as:

$$V = 2\pi^{3/2} w_x w_y w_z \cong \left(2\pi w_x w_y\right)^{3/2}.$$

Here, the volume was estimated from two measured radii instead of three. To verify that the shape of the cloud is indeed symmetric and the approximation above is valid the atomic fluorescence was detected with an additional CCD camera oriented at 45 degree with respect to the imaging beam.

Finally, the number of atoms versus time was fitted by (5) where $\sqrt{2\pi}^3 w^3$ was replaced by $V$, which was averaged over the first 2 seconds of MOT decay. While fitting, the linear loss rate $\gamma$ was assumed to be constant and equal to $0.12 \text{ s}^{-1}$, and the binary loss rate constant $\beta$ and $N_0$ were the fitting parameters.

## APPENDIX B: FREQUENCY INSTABILITY OF LASERS

The 530.7 nm laser was locked to the ULE cavity. Even though the cavity is quite stable, instability of the Green MOT AOM (in Figure 2) frequency leaded to a 100 kHz short range instability measured by our spectrum analyzer (HMS3010). This instability was the main source of statistical error.

Additionally, the cavity frequency slowly drifted, hence periodically verifying the detuning of the laser with respect to the cooling transition was required. This was accomplished by observing the MOT behavior while varying the detuning of the laser frequency. When approaching resonant frequency, the MOT cloud expanded quite rapidly with detuning. In order to observe this fast expansion, the MOT was loaded at detuning $\delta_{530} = -4\Gamma/(2\pi)$ and then detuning was rapidly changed to the target value. The target value was scanned to find a frequency at which cloud started to quickly expand, which was treated as the zero detuning. To verify this method and understand the error bar on determination of zero detuning, we performed direct spectroscopy of the 530.7 nm transition in the a MOT working on the strong 410.6 nm transition [52]. Both methods agreed on line position within 100 kHz, therefore 100 kHz was taken as our frequency uncertainty.

## APPENDIX C: TIMING OF THE DECAY EXPERIMENT

In our measurement of the MOT decay, we used the pulse sequence of 5 digital and 2 analog channels presented on the Figure 7. Digital channels controlled the laser beam for the ZS (ZS), cooling beams of the molasses (Molasses 410 nm), a trigger for the CCD camera that was taking photos of the cloud (Camera exposure), and the imaging beam AOM (Probe AOM) and mechanical shutters (Probe Shutter). Shutters were used to exclude any influence of AOM leakage on the MOT dynamics. Two analog channels changed the detuning of the MOT beams ($\delta_{530}$) and their power (MOT).

# APPENDIX D: SYSTEMATIC ERRORS

There were several sources of systematic errors. One of them was the laser frequency uncertainty and was discussed above. Others were the uncertainty in the number of atoms and cloud volume and the error of the fitting procedure. They are explained below.

We found the number of atoms in the MOT and its volume from an analysis of the absorption images. First, the CCD camera was calibrated with a laser beam of known intensity. A beam diameter was measured with accuracy better than 1% via a knife-edge method [53,54]. Power $P$ of the laser beam was measured with a power meter from Thorlabs (with a sensor S121C) which was last calibrated on 03.2014. We assume this calibration was off by no more than $\delta P / P = 15\%$, since this was the maximum discrepancy between different calibrated devices. This uncertainty in the laser power transforms into an uncertainty of $\frac{s}{1+s}\frac{\delta P}{P} < 4\%$ in the number of atom since the saturation parameter $s \approx 0.3$ for the imaging beam.

By comparison of measured beam profile by knife-edge method with one obtained by the CCD camera, we calibrated the size of a pixel and its sensitivity. The size of the pixel matched the camera specification. In a similar way, we measured an exact magnification of the imaging system. Overall, the error bar on linear geometrical dimension did not exceed 2% in comparison of two measurements.

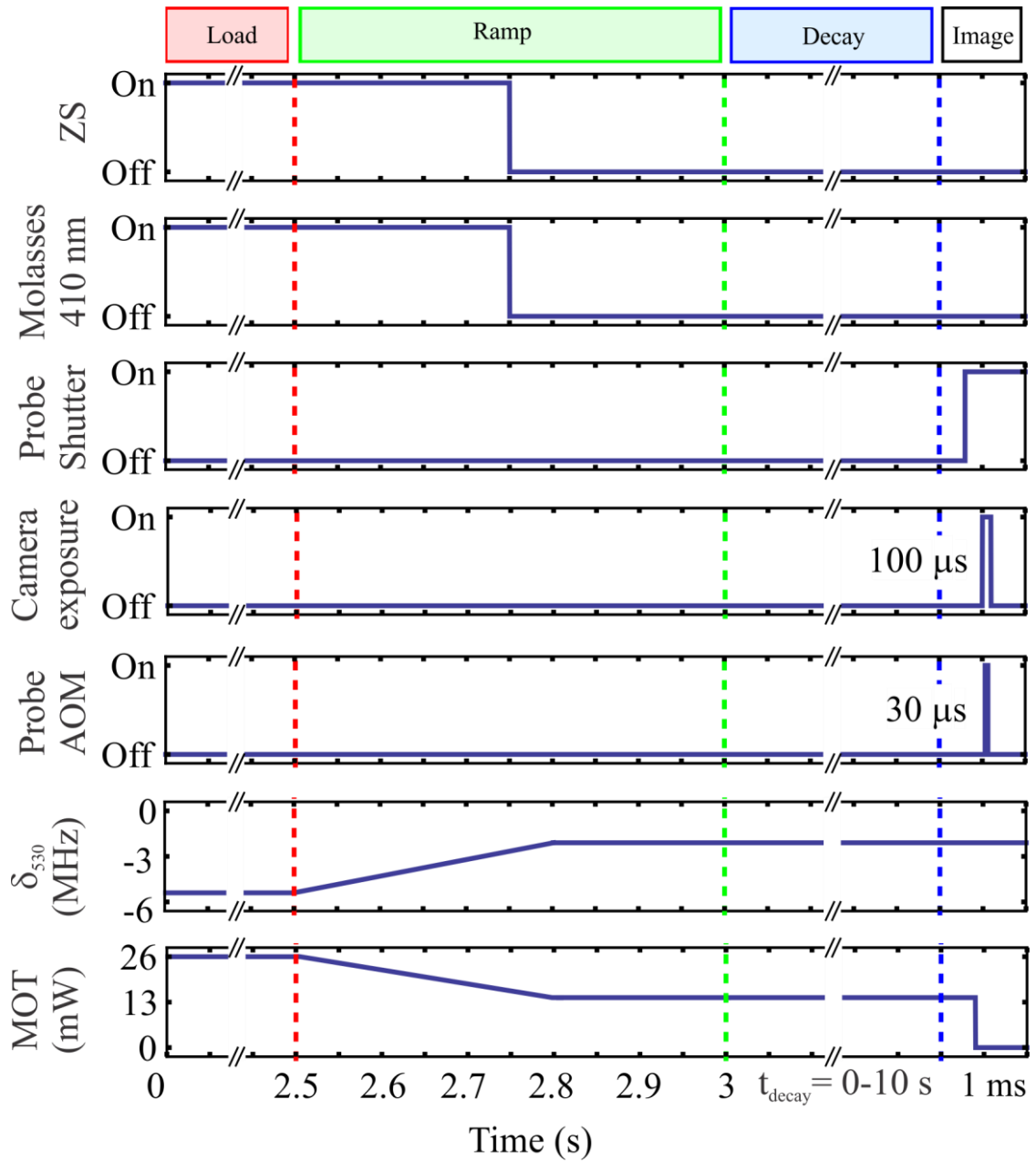

Figure 7 Pulse scheme used in the experiment. MOT power (MOT) is indicated as power per beam

In all our measurements, the MOT cloud was approximated by the Gaussian profile, which should be a correct model for the cloud in a harmonic potential. Nevertheless, due to imperfections of our setup, the profile of the cloud was not exactly Gaussian, therefore leading to uncertainty in the measured number of atoms and the size of the trap. To estimate this uncertainty, we found the number of atoms and the cloud volume by numerical integration of the image. In calculating error bar for the volume, the square of the density of the atoms was computed. This measurement was done only for the first 2 seconds of decay to minimize errors in numerical integration, which did rise with decay time. The source of this noise is a frame taken with no atoms, which is a bright frame in the case of absorption imaging, so it had considerable absolute level of noise due to the laser power fluctuations. Subtraction of frame with and without atoms therefore introduced considerable noise, especially on wings of the atomic distribution.

Note, that the uncertainty on a computed integral of the square of density is in fact gives error in the binary loss rate constant, which was found to not exceed 25%. The atom number measured via numerical integration and the one found assuming a Gaussian profile were different by no more than 20%. The volume uncertainty was less than 25%.

Additional sources of possible error include uncertainty in the polarization of the imaging beam. During imaging, the signal collected by the CCD camera also depends on a polarization of the imaging light. For example, in the case of the circular polarization used in our experiments, all atoms will be pumped on one of the states with a maximum possible component of total angular momentum leading to effective saturation parameter 50% different from the case of linear polarization. To calculate the number of atom in the MOT we did take polarization into account by computing the efficient saturation parameter for the polarization we used. We verified agreement between our theoretical model and measured quantities by comparing images taken with the same parameters of the MOT but with different imaging beam polarization. At these conditions, the number of atoms was kept the same, but imaging signal was different due to the discussed polarization effect. These images were in good agreement (within 3%) with the theory we used.

Our overall systematic uncertainty in the binary collision rate constant did not exceed 35%.